\documentstyle[11pt,paspconf,epsf]{article}

\begin{document}

\title{Population Gradients in Galaxy Clusters at 0.2 $<$ Z $<$ 0.6}

\author{E. Ellingson}
\affil{Center for Astrophysics and Space Astronomy, Univ. of Colorado, Boulder, CO, 80309}

\author{H. Lin}
\affil{Steward Observatory, Univ. of Arizona, Tucson, AZ, 85721}

\author{ H.K.C. Yee and R. G. Carlberg}
\affil{Dept. of Astronomy, Univ. of Toronto, Toronto, Ontario, M5S 3H8, Canada}

\begin{abstract}
We present a principal component analysis of galaxy spectra 
from the CNOC sample of rich X-ray luminous clusters at $0.18 < z < 0.55$.
Composite radial distributions of different stellar populations
show strong gradients as a function of cluster-centric redshift.
The composite population is dominated by evolved
populations in the core, and gradually changes to one which
is similar to coeval field galaxies at radii greater than the virial
radius. We do not see evidence in the clusters for an excess of star formation
over that seen in the coeval field.  Within this redshift range, significant
evolution in the gradient shape is seen, with higher redshift
clusters showing steeper gradients. This results in larger
numbers of younger galaxies seen towards the inner regions of the
clusters-- in effect, a restatement of the Butcher-Oemler
effect. Luminosity density profiles are consistent with
a scenario where this phenomenon is due to a decline over
time in the infall rate of field galaxies into clusters. Depending on
how long galaxies reside in clusters before their star formation
rates are diminished, this suggests an epoch for maximal infall
into clusters at $ z > 0.7$. We also discuss alternative scenarios
for the evolution of cluster populations.

\end{abstract}


\section{Introduction}

The evolving populations in galaxy clusters
offer a unique opportunity to observe galaxy
evolution in action, and particularly the 
effects of environment on star forming galaxies.
Present-day rich clusters have strikingly different
populations from galaxies in poorer environments,
suggesting that some mechanism is at work 
transforming normal field galaxies into the cluster
population.
Many recent investigations have focused on the
details of how this tranformation occurs (Couch \& Sharples 1987,
Barger et al., 1996, Poggianti et al., 1999, Balogh et al., 1999).
The emerging
picture is that there may be a population of galaxies
which were formed very early in the cluster's history,
corresponding to the ellipticals often seen in cluster cores.
(e.g. Bower et al., 1992, Ellis et al., 1997)
Subsequent generations of infalling field galaxies have
had their star formation disrupted, possibly with 
an associated starburst. As this transformation
progresses, these galaxies might be identified with normal-looking
spirals, then galaxies with strong Balmer absorption spectra,
and finally S0 galaxies which have retained some of their disk structure
but have ceased active star formation.

Most of these insights have been obtained
by observing distant clusters,
as higher redshift clusters have been shown
to contain a higher fraction of star-forming galaxies--
first described as the Butcher-Oemler effect
(Butcher \& Oemler 1984).
Ostensibly, the processes which create present day
clusters are at an earlier stage of their work at these
epochs, and these clusters provide a time sequence
for observing the growth of the cluster structure. 
Here we describe an investigation into the relationship
between galaxy evolution and cluster structure, based
on a well-defined  and fairly homogeneous sample of
intermediate redshift clusters, and coeval field
galaxy measurements. In an approach complementary
to that of looking at galaxy properties in detail,
we instead focus on building smooth composite
spatial distributions within the cluster of the various stellar 
populations. With these distributions, it is
possible to chart the relationship between the evolution of galaxy populations
and the growth and evolution of the cluster.

\section{The CNOC Cluster Sample}

The CNOC (Canadian Network for Observational Cosmology)
Cluster Redshift Survey targeted 16 rich
X-ray luminous galaxy clusters with $0.17 < z < 0.55$.
Deep Gunn $g$ and $r$ imaging and multi-slit spectroscopic observations
from the Canada-France-Hawaii 3.6m telescope were used to map
the cluster sample to radii of 1--3 h$^{-1}$ Mpc
from the cluster cores (see Yee et al., 1996).
Wavelength coverage was $\sim 3500-4500$ \AA ~in the rest frame
of each cluster, with resolution of about 15 \AA.
A total of 1200 cluster galaxies were spectroscopically identified.
Particular care was taken to quantify
selection effects and the completeness of the sample
as a function of galaxy magnitude, color, redshift and position.
The wide field coverage and careful attention to
the empirical selection functions are crucial for 
building an accurate  portrait of cluster structure.

Dynamical and spatial analyses of the clusters
(Carlberg et al., 1996, Carlberg, Yee \& Ellingson 1997,
Carlberg et al., 1997a,b ) yielded cluster masses, mass-to-light ratios
and density profiles. Lewis et al., (1999) presented
the X-ray gas profiles from ROSAT HRI and PSPC observations
of much of the sample. Balogh et al., (1997, 1999) analyzed
spectral line indices for cluster galaxies.
Here we combine measurements of
the stellar populations in the cluster galaxies with
the dynamical and X-ray properties of the clusters
to address the issue of galaxy evolution in terms 
of the spatial distribution of different populations. 

\section{Principal Component Analysis}

Principal component analysis (PCA) provides a sensitive method
for measuring the strengths of different stellar
populations from spectroscopic data.
This technique has been used by other groups (e.g. Connolly et al.,
1995, see also Zaritsky, Zabludoff \& Willick 1995) to
determine galaxy populations.
It is especially applicable to the
fairly noisy high-redshift data of the CNOC survey,
because the entire spectrum, rather than
a narrowly defined range of line indices, contributes to the measurement.
We project each spectrum as the sum of
four orthonormal galaxy ``vectors." These
vectors are derived from relatively high quality galaxy
spectra in the Las Campanas Redshift Survey
(Shectman et al., 1996) at low redshift. Figure 1 shows the four
original vectors, which correspond roughly to two ``Emission line"
components ([OII] and [OIII], which we combine to a single vector
because our spectra do not extend to the H-$\beta$/[OIII]
region), a ``Balmer component" featuring
strong Balmer absorption lines indicative of intermediate
age stars, and an ``Elliptical"
spectrum showing the standard old-population features.
\begin{figure}
\plotone{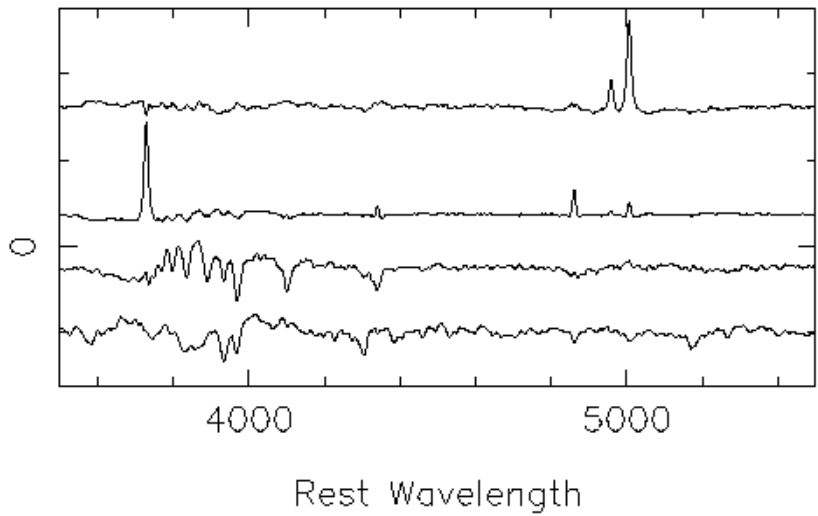}
\caption{Principal Component Vectors derived from the Las 
Campanas Redshift Survey. Four vectors are defined: two which are combined to
form our ``Emission" vector, ``Balmer," and ``Elliptical".}
\end{figure}
The power in each of the three components 
provides a measurement of the stellar population.

The combination of these three PCA components are adequate to
separate most types of normal galaxies. Figure 2 shows PCA
components for galaxies in the Kennicutt Spectral Atlas of
low redshift galaxies (Kennicutt, 1992). Normal elliptical
and spiral galaxies are well separated on this plot.
\begin{figure}
\plotfiddle{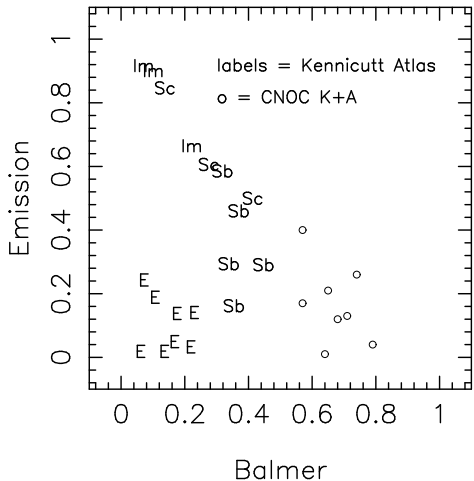}{2.5 truein}{0.0}{50}{50}{-150}{-20}
\caption{PCA Decomposition of Galaxies from the Kennicutt 
Spectral Atlas. Morphological types of the low redshift galaxies are 
labeled. Open circles denote K$+$A galaxies from the CNOC clusters.
The normalization of the sum of the three
components to 1 means that non-zero data occupy a triangular region
on this plot. }
\end{figure}
Note that the Kennicutt Atlas does not populate the region 
of the plot with large Balmer values and low emission line values.
In this region we plot high signal-to-noise ratio examples of K$+$A galaxies
from the CNOC clusters. 


\section{Population Gradients in Galaxy Clusters}

A composite average population gradient
from  15 clusters was constructed
from galaxies with k-corrected $M_r < -19$ (Ho=100 km s$^{-1}$ Mpc$^{-1}$,
qo=0.1). 
Cluster membership was delineated  by a redshift within
3 times the line-of-sight velocity dispersion
from the mean redshift.
Galaxy positions were scaled by a
characteristic radius, $r_{200}$, determined for each cluster
by Carlberg et al., (1996), to normalize for
differing cluster richness, and 
so that values at a given $r/r_{200}$ represent those at
a fixed overdensity. Typically, $r_{200} \sim$ 0.75--1.5 h$^{-1}$
Mpc for these clusters, and virial radii are 1.5-2 r$_{200}$. 
These dynamical parameters are well
determined for these clusters by 30-200 cluster members
each, and have been found to yield results consistent with
X-ray determinations of the cluster masses (Lewis et al., 1999).
The CNOC cluster MS0906+11 was omitted because
it shows strong binary substructure in both X-ray and
velocity data, and robust values for dynamical parameters
were not possible.

Weights for each galaxy were calculated as inverses of the
completeness functions for magnitude, color and
position within each cluster field  based on
deep two-color galaxy photometry in the fields
(see Yee et al., 1996 for details
on completeness functions). Using weights such as these
ensures that our spectroscopic sample yields a
result representative  of the entire cluster.

A clear spatial gradient is seen, with the elliptical-like
population declining from the cluster core towards the outer
parts of the cluster, and an accompanying increase
in emission line galaxies (Figure 3).
\begin{figure}
\plotfiddle{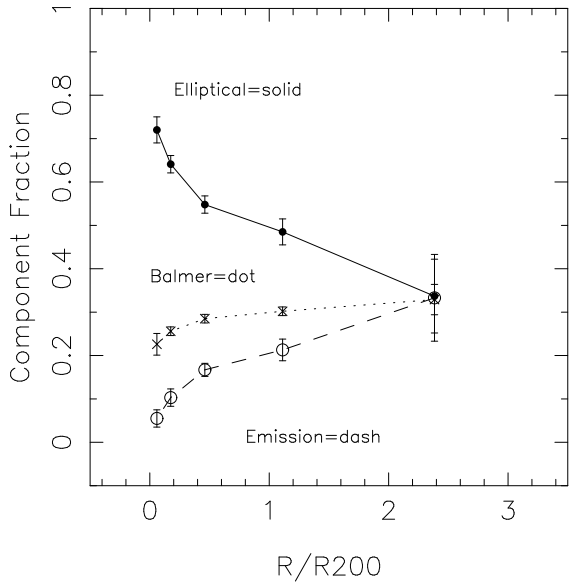}{3.0 truein}{0.0}{50}{50}{-200}{-50}
\caption{Composite radial gradients from 15 clusters based on the
vectors shown in Figure 1. Normal field galaxies at these redshifts
have values approximately equal to [0.30, 0.40, 0.30] for
Elliptical, Emission and Balmer, respectively.} 
\end{figure}
There are between 20 and 294 galaxies in each bin;
uncertainties were estimated via bootstrap techniques.
A self-consistent correction for projection effects was
tested, assuming an average spherical galaxy distribution, but is not
included in the figures shown here.  This correction
tends to steepen
the gradient very slightly but is very small,
generally raising the innermost point for the older
populations  just a few percent.

Morphological and spectral gradients similar
to this are seen in low-z clusters (e.g. Oemler, 1974; Dressler, 1980; Whitmore
et al., 1993).
Similar gradients over large scales at higher redshifts  have also been 
noted by Abraham  et al.,  (1996),
Couch et al., (1998) and Dressler et al.,  (1997). 
Balogh et al., (1997) traced the star formation gradients in this
CNOC sample via measurements of the [OII] emission line and
also found a similar trend.
Such gradients are
consistent with
the infall of field galaxies in hierarchical clustering
models of cluster growth (e.g. Gunn \& Gott, 1972; Kauffmann et al.,
1995a,b) and suggest that galaxies which have more recently
fallen into the cluster potential preferentially inhabit its
outskirts.

The PCA components can be transformed to
measure a galaxy's match to any specific stellar population.
\begin{figure}
\plotfiddle{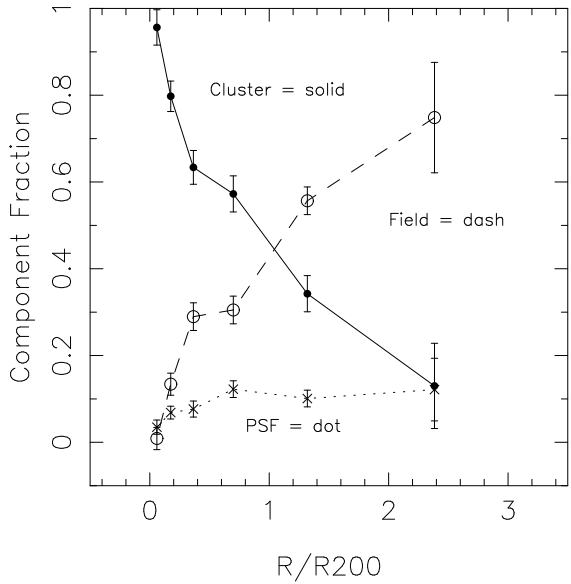}{3.0 truein}{0.0}{50}{50}{-200}{-50}
\caption{ Composite radial gradients in transformed coordinates
(see text).}
\end{figure}
In Figure 4 we plot a weighted composite for the sample, 
where
the components are redefined. ``Cluster" values were
chosen to match ellipticals in the Kennicutt Atlas (1992),
and agree with the reddest galaxies in these clusters.
Spectral synthesis models (Charlot \& Bruzual, 1993)
indicate that any population
more than about 3 Gyr old will yield similar PCA values.
The new ``Field-like" vector is calculated separately for each cluster,
based on the average for field galaxies
at these same magnitudes at the cluster redshift from
the CNOC2 Field Redshift Survey 
(Carlberg et al., 1998).
On average, values are similar to a present-day Sbc, but with
additional power in Balmer absorption lines. There is a mild
increase in the ``Emission" fraction between z=0.2 and z=0.6. 
``Post-Star-Formation (PSF)" values are determined 
from a spectral synthesis model of a galaxy whose star formation
is constant for 4 Gyr and then is abruptly stopped.
The new vector corresponds to the time of the maximum Balmer component,
about 0.5 Gyr after the truncation,
and is in general agreement with PCA values from
the strong K$+$A spectra of galaxies seen in the clusters.

Note that these components are not strictly orthonormal, though they
are normalized to a sum of unity, as before.  Thus, the equivalent of total
flux is conserved, but the component values are dependent on the
choice of all three vectors. While the ``Cluster" and
``Field-like" components are robustly determined empirically, the
``PSF" component is more arbitrarily defined. Choosing
a slightly different recipe for this component will raise or lower
all three values slightly.

Despite this potential degeneracy,
the overall cluster structure is well illustrated by
this transformation. Old population
galaxies preferentially occupy the cluster interior, and a
smooth gradient towards younger populations is seen until
the properties of cluster galaxies (defined from their redshifts) 
asymptotically approach those of the coeval field population
at 2--5 r$_{200}$, well outside of the virial radius.

The post-star-formation component appears relatively flat,
with a component fraction of about 10\%. This is larger than the fraction
of identifiable K$+$A galaxies seen in the same dataset by
Balogh et al., (1999), but smaller than the fractions
seen by Poggianti et al., (1999) in the MORPHS sample of clusters at similar
redshifts. The PCA technique, of course, is sensitive to
smaller fractions of galaxy light in the post star formation state,
so higher fractions are expected than for studies which
count only galaxies with extreme signatures.
Thus the comparison with Balogh et al., appears reasonable.
However, the comparison with the MORPHS clusters is more troubling, 
and may signify true differences in cluster populations
for X-ray selected clusters such as the EMSS sample,
versus the optically selected MORPHS clusters. One possibility is
that the former sample is likely to be
more relaxed and centrally concentrated, whereas optically selected
clusters may include objects with more recent infalling
structures, which may explain a higher fraction of galaxies
which have recently had their star formation truncated.

In Figure 5 we examine the trend in population gradients in clusters
over our redshift range of 0.18 to 0.55.
\begin{figure}
\plotfiddle{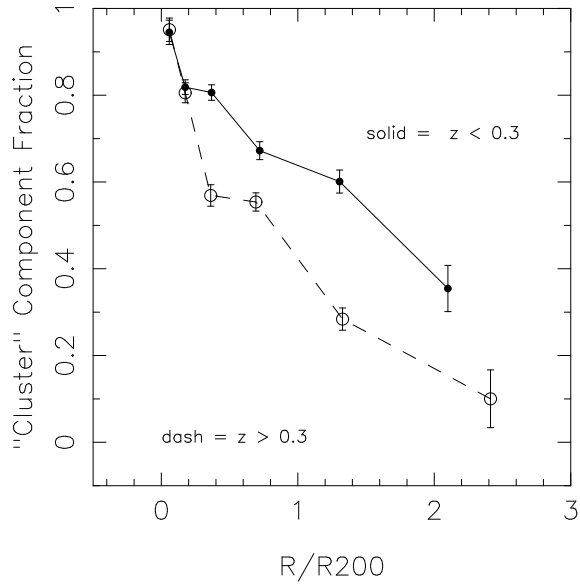}{3.0 truein}{0.0}{50}{50}{-200}{-50}
\caption{ Composite radial gradients in the ``Cluster" component
as a function of cluster redshift.}
\end{figure}
Plotted are the ``Cluster"
components versus radius for samples at $0.18 < z < 0.3$ and
$0.3 < z < 0.55$. The PCA values,
weights and uncertainties are
calculated as before. While the cluster cores
appear to be similarly dominated by old populations,
a significant change in the cluster gradient is seen 
between the two redshift bins. At lower redshift, the old
population appears to dominate to larger radii, whereas
at high redshifts, the field galaxies are more noticeable
in the inner portions of the cluster. 
This steepening of the population gradient
can be thought of as a more detailed restatement of the
Butcher-Oemler effect, which describes
the increasing blue fraction in higher redshift clusters,
typically measured inside of 0.5 h$^{-1}$ Mpc.


\section{Luminosity Density Profiles}

From the principal component analysis, relative luminosity 
surface density
profiles can also be constructed, to trace the total light
associated with the various stellar populations. Figure 6 
shows the profiles for the transformed vector coordinates
described above. 
\begin{figure}
\plotfiddle{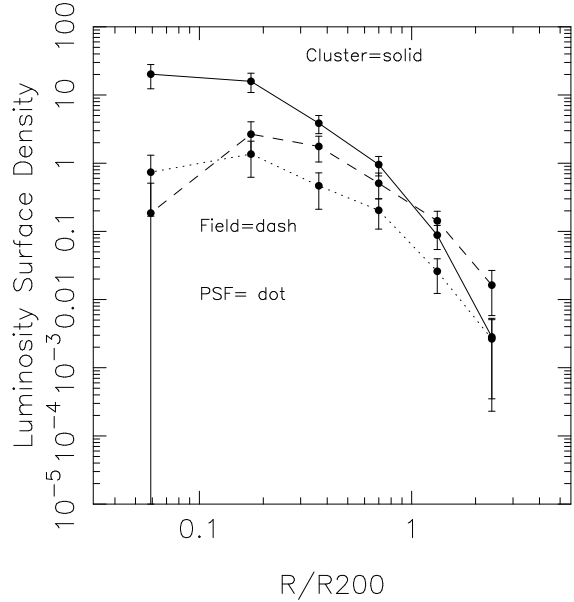}{3.0 truein}{0.0}{50}{50}{-150}{-10}
\caption{ Composite luminosity density profiles from 15 clusters
based on the transformed vector coordinates. The profiles show
an increasing spatial extent from ``Cluster" to ``Post-Star-Formation"
to ``Field-like."}
\end{figure}
The fractional gradients seen in Figures
3 and 4  are shown here to be caused by light distributions
which increase in extent from older to younger.
This is quite consistent with what is expected from
the accretion of field galaxies into the cluster. At first,
infalling galaxies should have a relatively flat distribution,
and as the galaxies become dynamically incorporated into
the cluster, their distribution will appear more
virialized, with a tighter profile. This dynamical
evolution occurs in tandem with spectral evolution,
causing the observed population gradients.

In Figure 7 we illustrate a possible explanation for the
evolution in cluster populations and population gradients
in clusters by constructing luminosity profiles
as a function of cluster redshift.
Figure 7a shows the relative
luminosity surface density for the old ``Cluster" component for
$0.18 < z < 0.3$ and $0.3 < z < 0.55$. The two curves
are normalized to have the same total number of galaxies
with 0.5 h$^{-1}$ Mpc in both bins. It appears that the old population
has not changed in shape significantly over this timescale,
which is in agreement with the idea that these galaxies are in
a stable dynamical equilibrium. This population contains
light both from  any elliptical galaxies which may have
formed early in the cluster history, as well as once star forming  galaxies
which have had their star formation truncated 3 Gyr or more ago.

Figure 7b shows the sum of the ``Field-like" and ``Post-Star-Formation" 
light profiles, representing galaxies which are newer introductions
to the cluster potential. 
These curves are normalized identically
to those in Figure 7a. Here the Butcher-Oemler effect is again
obvious, with a higher density of younger populations being
seen at higher redshifts. Outside the core region ($>$ 0.25 r$_{200}$)
these curves also appear
to be similar in shape, suggesting that dynamical state (i.e.
time since entry to the cluster potential) and age of the
stellar populations are fundamentally linked. This is a reasonable
assumption if the mechanism which transformed the population
is due to the cluster environment.
The main difference in these curves is then a simple vertical
shift, and these results can thus
be interpreted as a decline in the infall rate of
new galaxies into the cluster.  
\begin{figure}
\plotfiddle{ellingsone_7a.eps}{1.0 truein}{0.0}{40}{40}{-210}{-100}
\end{figure}
\begin{figure}
\plotfiddle{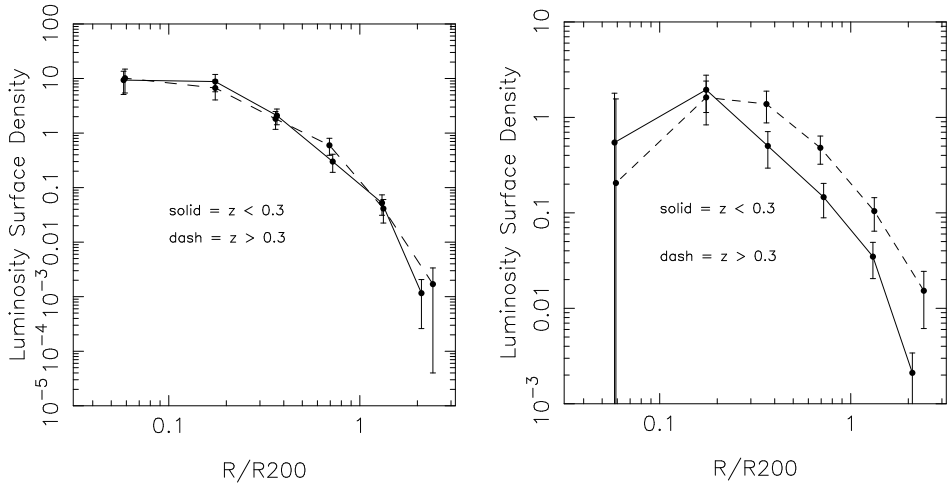}{1.0 truein}{0.0}{40}{40}{-25}{-3}
\caption{ Composite luminosity density profiles from 15 clusters
as a function of redshift.
Figure 7a (left) shows profiles for the ``Cluster" component for
clusters with $0.18 < z < 0.30$ (solid) and $0.30 < z < 0.55$ (dash).
Figure 7b (right) shows the
profiles for the sum of the ``Field-like" and  ``PSF" components and
the same redshift bins.}
\end{figure}

Linking this interpretation to cosmological models remains
uncertain because of the unknown timescales of both virialization
and population evolution in these environments. For example, 
the mechanism which affects ongoing star formation in clusters
may operate quite gradually, and the ``field" galaxies we see in
the cluster may have been introduced to the cluster a billion
or more years earlier than the observed epoch.
The observed
epochs are at $z \sim 0.23$ and $ z \sim 0.43$;
a rough estimate of at least 1 billion year delay
between infall and the 
start of the spectral transformation 
would place the epoch
of maximal infall into clusters at $z > 0.7$, 
in agreement with predictions from a low-density universe.

\section{Other Scenarios for Galaxy Evolution in Clusters}

An alternate viewpoint to a declining infall rate is that
some physical  property of the cluster environs has
evolved over these timescales, driving the observed evolution
in the populations. Possibilities include 
evolution of the infalling field population, 
differing rates or efficiencies of galaxy-galaxy interactions, 
and evolving gas densities in the cluster, which would affect the efficiency
of ram-pressure stripping of gas from infalling galaxies.
We have essentially removed the effects
of field galaxy evolution by constructing vector components
which are based on the observed field population at the
same magnitudes and redshifts as the clusters.

Global evolution in the X-ray gas is also probably not a strong
driving force in the observed evolution. All of the clusters
in our sample are luminous X-ray clusters, and there does
not appear to be a strong correlation between the X-ray
gas and the galaxy populations. 
\begin{figure}
\plotfiddle{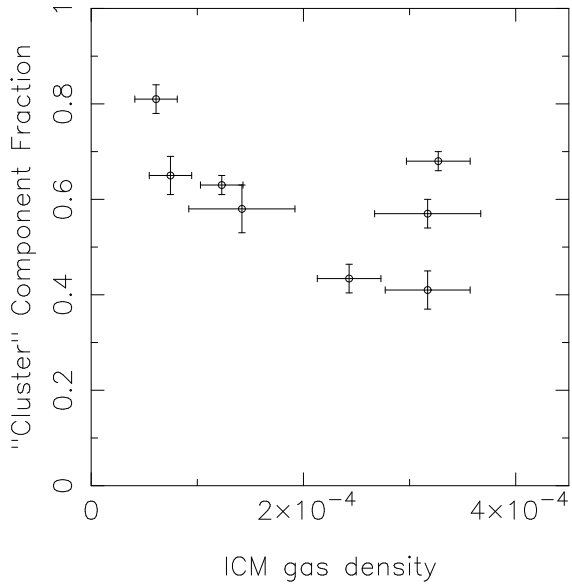}{3.0 truein}{0.0}{50}{50}{-180}{-50}
\caption{ ``Cluster" component value versus X-ray gas density,
(units of cm$^{-3}$) as measured at 0.5 r$_{200}$.
 direct link between cluster population evolution and X-ray
gas evolution would predict a positive correlation.
None is seen, suggesting that other factors drive the
observed evolution. } 
\end{figure}
Figure 8  shows
the old ``Cluster" component for 8 clusters versus
azimuthally averaged X-ray gas density, each measured at 
0.5 r$_{200}$ from the cluster core. While there is some
variation in each parameter, they do not appear positively
correlated, as would be expected if gas density evolution
were driving the overall population changes. (Note that gas stripping
may still be responsible for the transformation of individual galaxies,
while a cosmologically changing condition drives the population
evolution.)

Another possibility is that at higher redshift, infalling
galaxies have a larger reservoir of interstellar gas, and 
hence are more resilient to stripping. If this were the case,
we would expect to see that the ``Field-like" population
would have a tighter, more evolved 
spatial distribution at higher redshift than at lower redshifts.
Figure 7 cannot quantitatively rule out this possibility; however
the data appear to be more consistent with the distributions
keeping the same shape, with a simple  vertical shift, rather than
the higher redshift distributions having a higher peak towards
the inner regions due to a dynamically relaxed but spectroscopically
unevolved population.

Galaxy-galaxy interactions remain a possible mechanism for
galaxy evolution in clusters, although morphological evidence
for this remains mixed (e.g. Oemler et al., 1997). Since
the effects of cluster richness and dynamics are normalized
in this analysis, any evolutionary effect must come from either
a cosmological change in the clustering of galaxies as they
fall into the cluster (i.e. a declining rate of
infalling pairs or small groups which might preferentially
inhabit the regions near clusters), or in the resiliency of 
galaxies at continuing to form stars in the wake of a collision. 

\section{Conclusions}

We have carried out a principal component analysis of 
galaxies in 15 clusters at intermediate redshift to
investigate the relationship between cluster population and
global spatial/dynamical structure.  Composite population
gradients show a smooth transition from the
infalling field galaxy population to the older populations
seen in the cluster core. 
The gradients are corrected for the evolution of the field
population and cluster richness.
Evolution between z=0.6 and 0.2
is manifest in a flattening of this gradient at later epochs,
consistent with the Butcher-Oemler effect.
This phenomenon is most consistent with scenarios where
the mechanism which truncates star formation in individual galaxies
remains constant, but the cluster population evolution is
driven by a declining rate of infall into the
clusters. Scenarios in which galaxies at earlier times are
more resilient to  destructive environmental effects of the cluster
may also be possible.

\acknowledgments
EE acknowledges support provided by the National Science Foundation
grant AST 9617145.
HL acknowledges support provided by NASA through Hubble Fellowship grant
  \#HF-01110.01-98A awarded by the Space Telescope Science Institute, which 
  is operated by the Association of Universities for Research in Astronomy, 
  Inc., for NASA under contract NAS 5-26555.

\end{document}